\documentclass[conference]{IEEEtran}
\IEEEoverridecommandlockouts
\usepackage{cite}
\usepackage{amsmath,amssymb,amsfonts}
\usepackage{algorithmic}
\usepackage{graphicx}
\usepackage{textcomp}
\usepackage{xcolor}
\usepackage{booktabs}
\usepackage{url}
\def\BibTeX{{\rm B\kern-.05em{\sc i\kern-.025em b}\kern-.08em
    T\kern-.1667em\lower.7ex\hbox{E}\kern-.125emX}}
\begin{document}

\title{Extracting and categorising the reactions to COVID-19 by the South African public - A social media study
}

\author{\IEEEauthorblockN{Vukosi Marivate}
\IEEEauthorblockA{\textit{Department of Computer Science} \\
\textit{University of Pretoria}\\
Pretoria, South Africa \\
vukosi.marivate@cs.up.ac.za}
\and
\IEEEauthorblockN{Avashlin Moodley}
\IEEEauthorblockA{\textit{Department of Computer Science} \\
\textit{University of Pretoria, CSIR}\\
Pretoria, South Africa \\
avashlin@gmail.com}
\and
\IEEEauthorblockN{Athandiwe Saba}
\IEEEauthorblockA{ \\
\textit{Mail and Guardian}\\
Johannesburg, South Africa \\
athandiwes@mg.co.za}
}

\maketitle

\begin{abstract}
Social Media can be used to extract discussion topics during a  disaster. With the COVID-19 pandemic impact on South Africa,  we need to understand how the law and regulation promulgated by the government in response to the pandemic contrasts with discussion topics social media users have been engaging in. In this work, we expand on traditional media analysis by using Social Media discussions driven by or directed to South African government officials.  We find topics that are similar as well as different in some cases. The findings can inform further study into social media during disaster settings in  South  Africa and beyond. This paper sets a framework for future analysis in understanding the opinions of the public during a pandemic and how these opinions can  be distilled [in a semi-automated approach] to inform government communication in the future.
\end{abstract}


\section{Introduction}
It is over a year since the first reported coronavirus disease (COVID-19) case in South Africa \cite{mg2020}. Additionally, the level of government engagement toward the South African public can be characterised as unprecedented, with a rate and frequency unlike anything that has been seen in recent years. As part of the guidelines outlined by the World Health Organization (WHO), governments in Africa tried to balance their responses to the health crisis by also responding to the public on social and economic impacts of the disease and disease mitigation strategies \cite{gcro2020}. There are a number of examples where research continues to be conducted on studying the pandemic spread and the level of this engagement to disseminate knowledge in the South African context  \cite{mbuvha2020data,arashi2020spatial,marivate2020use}. In our approach, we investigated the social narratives related to the public engagement prevalent on social media as a window into some of the concerns that the South Africa public might have during a specific period of the COVID-19 response.

In this work, we extract patterns of social media interactions between the South African government (through its representatives) and citizens (the public) to understand the evolution of this engagement over time. To do so, extracting the discussion patterns on social media and contextualising the limitations present are important, including the fact that  social media is not a representation of all parts of the South African society. South Africa has 64\% Internet penetration with 36\% of the population engaging with any type of social media \cite{kemp2021}. If we concentrate on Twitter, South Africa has just over 2.3 million users on this platform \cite{kemp2021}, out of an estimated total population of 58.8 million \cite{statistics2019mid}.

In our study, we used an empirical approach to extract topics from social media data for South Africa during a specific period. We analyse this data to answer the following questions:
\begin{itemize}
    \item What are the most discussed issues?
    \item How have the discussions changed over time in relation to the announced policy changes by the government?
    \item What topics are similar and dissimilar between official and public sources?
\end{itemize}

The rest of the paper is organised as follows. Firstly, we provide some background to the government communication during pandemics. Then, we present our data, followed by, an analysis on the extracted topics. Lastly, we discuss and conclude our findings, which is followed by our concluding remarks.

\section{Background - COVID-19 progression in South Africa and online discussion}
\label{sec:background}
The COVID-19 response by the South African government has been multi-modal. This meant that communication strategies were presented across a variety of different strategies to reach all members of the public including media briefings, announcements, and question and answer sessions. These strategies were implemented as government tried to bring citizens into their confidence and build public trust ~\cite{mma2020}. 

At the beginning of the COVID-19 pandemic in South Africa, the government pushed to make sure that the public was informed of the progression of the pandemic and that the public was informed about policy changes that were being debated or tabled that would have an impact on the lives of South Africans.

The 2020 Minister of Health, Dr. Zweli Mkhize, held daily briefings on several topics related to the Healthcare system’s response to COVID-19. The president, Cyril Ramaphosa, held briefings for the public before major changes such as: introducing and declaring a national state of disaster, announcing lockdown and lockdown levels etc. Ministers of other departments that were heavily affected by these policy changes also held frequent briefings that clarified policy decisions and they also answered questions from both the media and the public that were asked beforehand about the implications of these decisions  \cite{mma2020}. These types of engagements between the government and the public were novel for South Africa as both the level of engagement to the public and the rate/speed of these engagements were timely and topical.

However novel and timely these engagements were, the public still had reservations and a variety of opinions that were counter intuitive to the strategies that were put in place on the COVID19 pandemic and interventions by government and other sectors. These challenges meant that is a growing need to find better ways to understand these reactions and interactions between the government and the public. Social media analysis and traditional media analysis can be useful in extracting topics that are important in tracking a pandemic and other public healthcare related issues  \cite{ghosh2017temporal}. Analysis of media information, whether it is about formally communicated information and/or social public conversation, can be extracted and used to help scientists, governments and the public highlight gaps in the engagement and highlight areas for improvement \cite{mma2020,ghosh2017temporal}. Social Media, in South Africa and beyond, has been used to engage in debate and organise groups that mobilise volunteers that assist the vulnerable \cite{volunteer2020} or for COVID-19 data collection efforts \cite{marivate2020use}, mirroring past use in times of disaster~\cite{chae2014public}.

\subsection{Related work}

There has been a flurry of work related to aspects of the COVID-19 response from governments and the interaction between government and the public since the onset of the Pandemic. For public health, a focus has been on perception surveys that are carried out to also assist in planning for roll outs of interventions – as this is a part of the community engagement strategies to manage public disasters by considering local problems within communities. In South Africa, an example of this are surveys carried out in collaboration with the Human Sciences Research Council (HSRC) among others on a regular basis with a variety of members including healthcare workers, members from the public, and business owners. A few examples of South African perception surveys are inquiries investigating the public understanding of COVID-19 \cite{SAMJ13049}, general vaccine hesitancy \cite{runciman2021uj}, job loss and its impact on people and government trust \cite{posel2021job} and health care worker knowledge on COVID-19 and the way forward \cite{manyaapelo2021determinants}. Unfortunately, limited to no work exists investigating social media and the interplay between public trust, misinformation, and engagement of the South African public. Internationally, there has been work exploring concepts such as world leader engagement on social media  \cite{rufai2020world}, public risk perception \cite{dyer2020public}, public attitudes in the USA and UK \cite{hussain2021artificial} and sentiment analysis \cite{lwin2020global}. Our work focused on extracting patterns from social media data covering South Africa in the first few months of the COVID-19 pandemic in 2020. We specifically used guided topic models to be able to identify the public discussions that had taken place during the period of study. Lastly, we focused on the public reaction to interventions of the government, so that further insight may be obtained.

\subsection{Limitations}
We collected Twitter data that is related to the South African COVID-19 response from the national government. Specifically, we stratified the information into national government leadership/health departments and provincial leadership/health departments. In our approach, we did not cover the full spectrum of reactions or opinion. Additionally, although it is not our focus for this paper to interrogate the full spectrum of responses prevalent on social media, we believe that a narrower more specific inquiry is needed to reduce the effects of noise in the data by contextualising this nuance. Furthermore, we focused on a few keywords and restrictions to South Africa to extract patterns that are useful for South Africa. If we were to explore the entire array of information during this time and incorporate a wide net of data collection, we would amplify the effects of bots and spam  \cite{kim2016garbage}. Further work can be done to identify bots in COVID-19 social media data in South Africa, especially after the rise of misinformation connected with COVID-19 \cite{nsoesie2020covid}. Lastly, we did not quantify the sentiment the public had in their discussions with official government sources. This is beyond the current scope of the paper, but we wish to explore this work in future to investigate a localised sentiment analysis approach.

\section{Methodology}
The problem identified in the early phases of this project was that an unsupervised topic model that operated on large data sets produced sub-optimal topics that do not support the analysis of public response on Twitter. To overcome this problem, a supervised and multi-tiered modelling approach was needed to guide the knowledge extraction to produce better quality topics for analysis. The Anchored Correlation Explanation (CorEx) \cite{gallagher2017anchored} was identified as a suitable candidate for this task. CorEx provides the functionality to provide the topic model with seed words that influence that topics produced. 

In our study, we had a focused data set consisting of official communication from government and we had a larger public response corpus that contained microblogs responding to key government officials on Twitter. We wanted to understand what the public were saying in response to official communication sent out by government officials on Twitter. To achieve this, our methodology consisted of the following steps:

\begin{enumerate}
    \item Keyword analysis on official Tweets
    \item Compilation of seed words
    \item Build official topic model leveraging seed words
    \item Extract top 10 keywords from each topic in the official topic model
    \item Leverage extracted keywords from the official model to seed a topic model on the public response data set
    \item Label official and public data with topics from their respective models
    \item Analysis of relationships between related topics from each model to understand the interaction between government and the public on COVID-19 related matter.
\end{enumerate}

\section{Data collection and exploratory analysis}

We collected data from Twitter that covers the periods of 01 March 2020 to 17 May 2020. We collected government communication from a few official accounts (see Table \ref{tab:accounts_tracked}). Traditional media analysis has highlighted that journalists had focused on President Ramaphosa and Dr Zweli Mkhize for comments when it comes to to COVID-19 in South Africa \cite{mma2020}. In our approach, we used  TWINT\footnote{\url{https://github.com/twintproject/twint}} to collect the data. The data collection was under ethical clearance \emph{EBIT/88/2020} of the Engineering and Built Environment Faculty of the University of Pretoria.

We collected 919,019 
social media posts from 189,266 unique users.
The data was segmented into two subsets, \textit{TweetsOfficial} and \textit{TweetsPublic}. \textit{TweetsOfficial} contains microblogs (twitter posts) published by accounts in Table~\ref{tab:accounts_tracked}, \textit{TweetsPublic} contains the remaining microblogs that reference the accounts in Table~\ref{tab:accounts_tracked}. \textit{TweetsOfficial} contains 2,537 microblogs from 4 unique users. 
and \textit{TweetsPublic} contains 807,044 microblogs from 189,262 unique users. 
\textit{TweetsOfficial} contains a coherent messages from government officials to the public whereas \textit{TweetsPublic} contains the general public responses to these and other information, which can be incoherent and noisy due to the number of users contributing to the data.

\begin{table}[t]
    \caption{Twitter keywords (in the form of accounts) tracked for our study}
    \label{tab:accounts_tracked}
\resizebox{\columnwidth}{!}{%
    \centering
    \begin{tabular}{ll}
    \toprule
     \textbf{National Account} & \textbf{Description}\\
 \midrule
@CyrilRamaphosa & President of RSA	\\
@DrZweliMkhize & Minister of Health, RSA\\
@HealthZA & National Department of Health [NDOH] \\
@nicd\_sa & National Institute of Communicable Diseases [NICD] RSA	\\
    \bottomrule
    \end{tabular}
}

\end{table}

\begin{figure}
    \centering
    \includegraphics[width=\columnwidth]{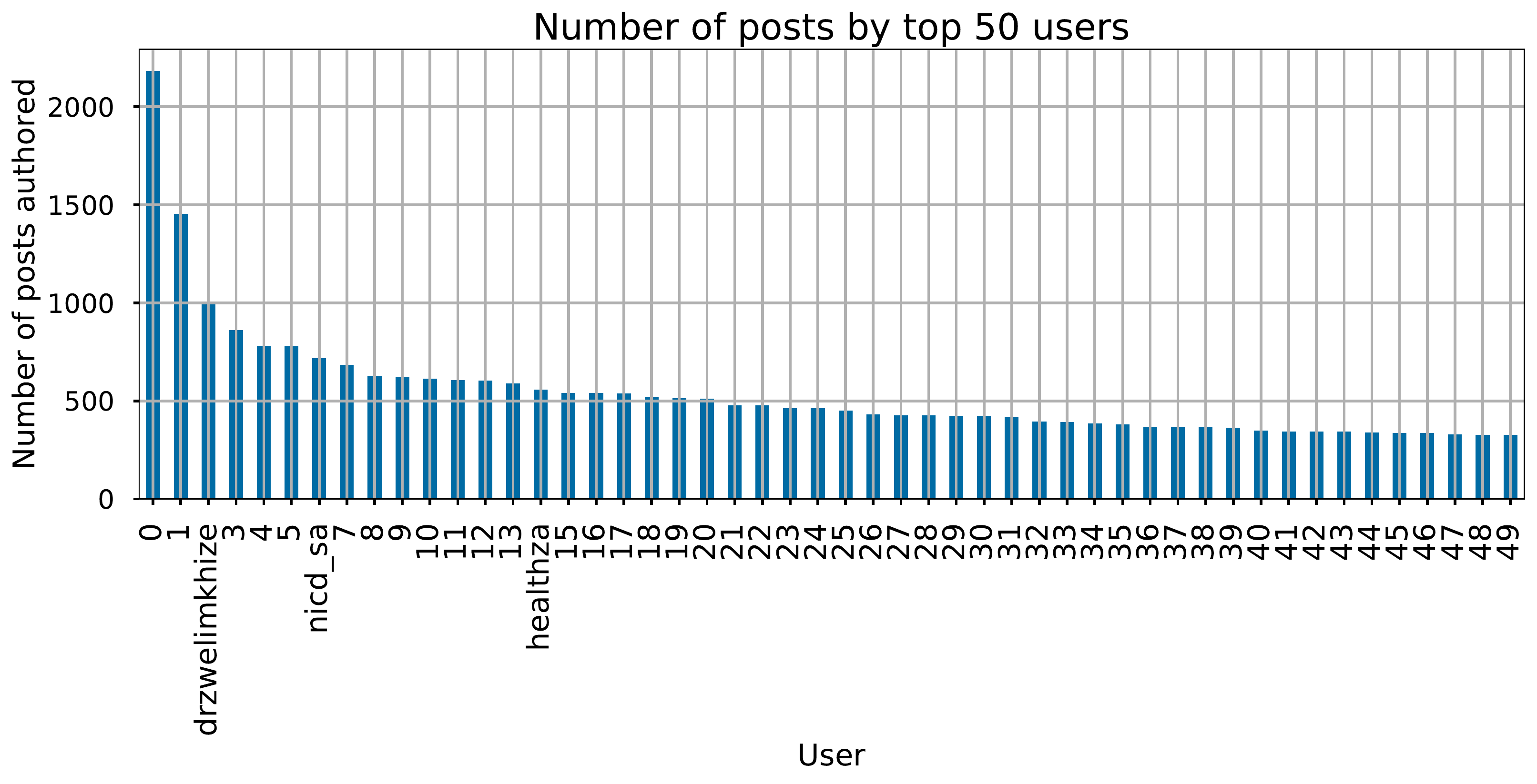}
    \caption{Top users and number of tweets (authored by each user) in the national dataset. }
    \label{fig:users_national}
\end{figure}

\begin{figure}
    \centering
    \includegraphics[width=\columnwidth]{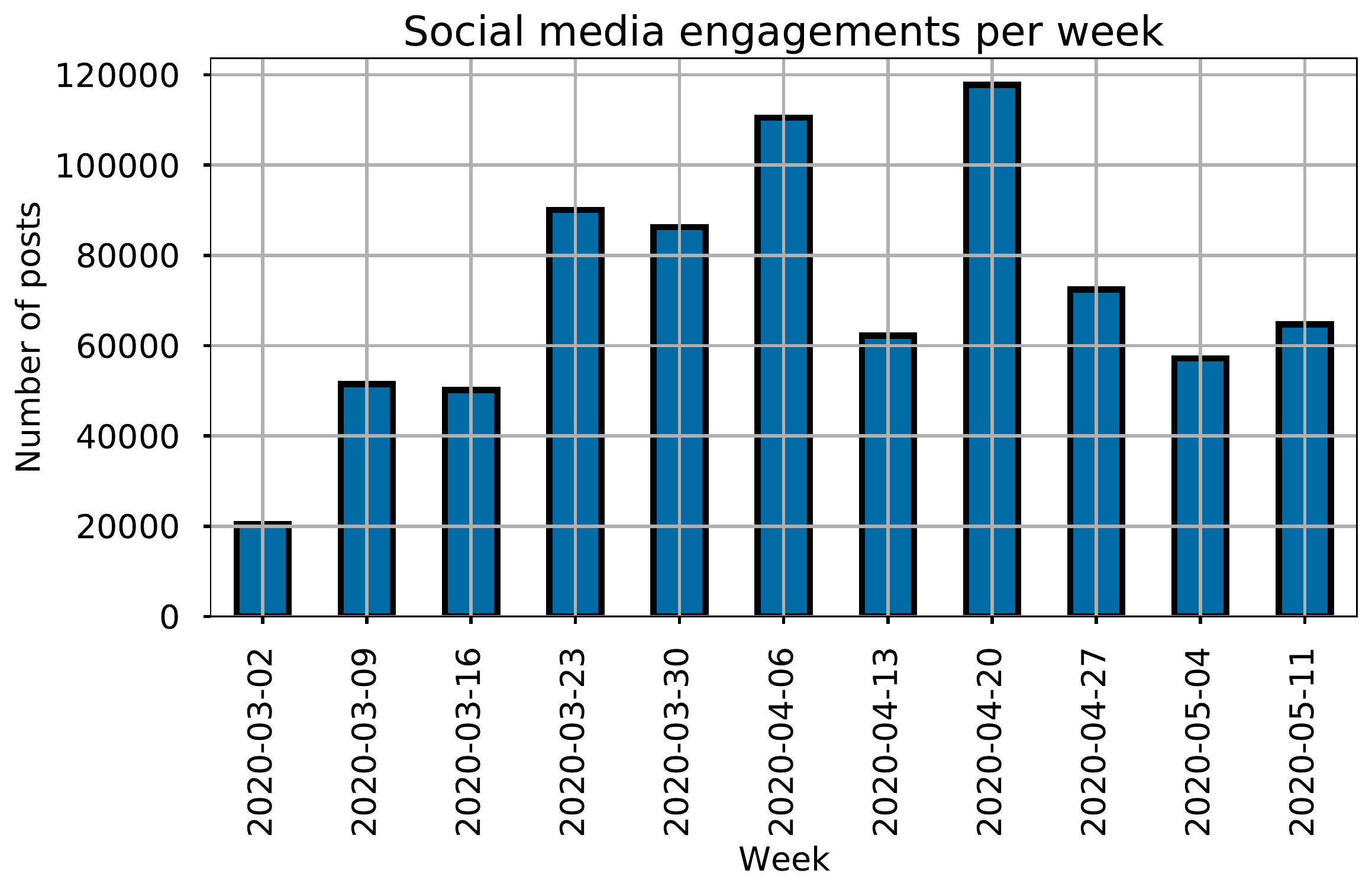}
    \caption{Weekly number of posts [Timeline]}
    \label{fig:users_weekly}
\end{figure}

\begin{figure}
    \centering
    \includegraphics[width=\columnwidth]{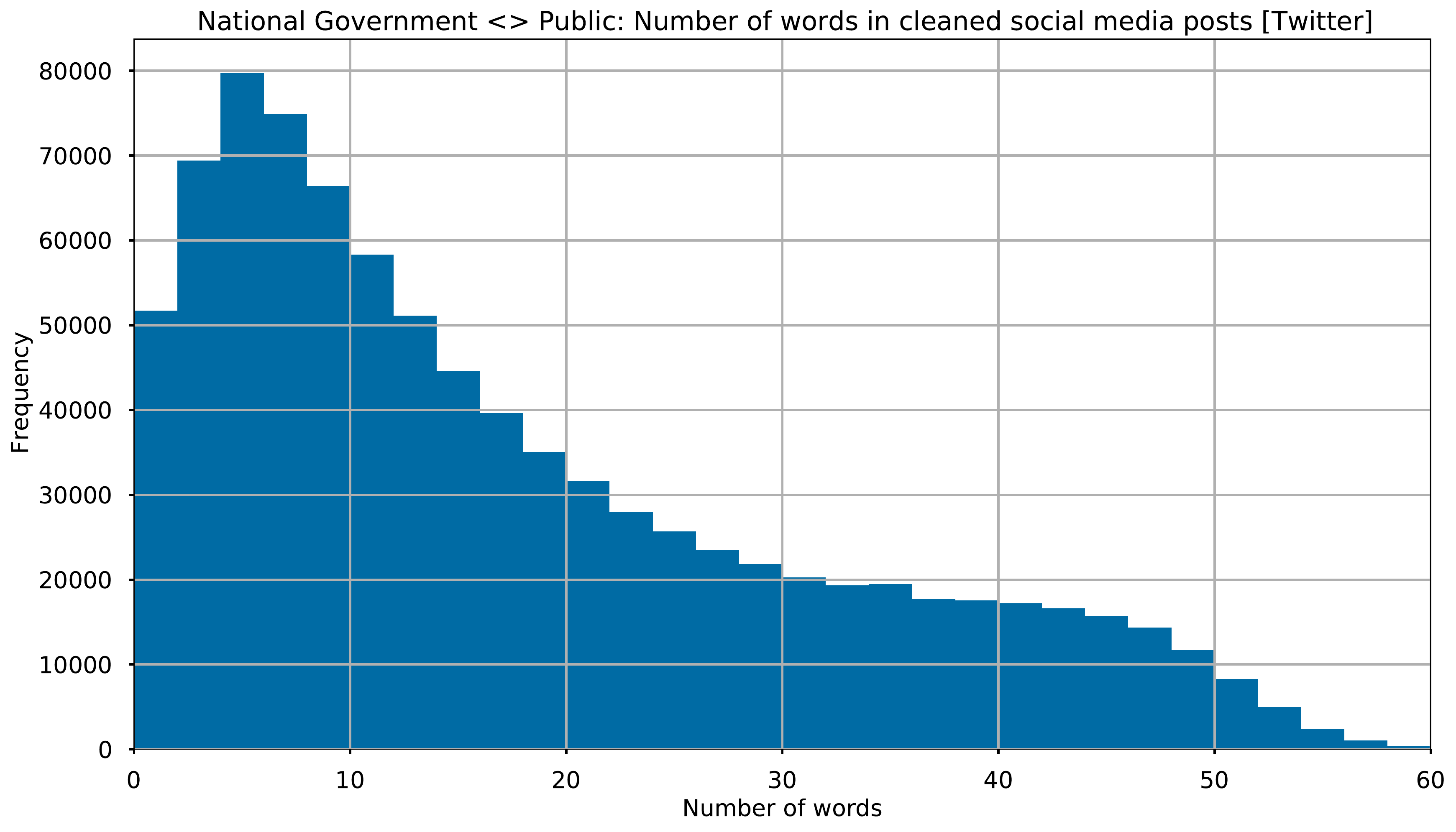}
    \caption{Number of words per social media post [Twitter post word frequency]}
    \label{fig:number_of_words}
\end{figure}

The number of posts by the top 50 users are shown in Fig \ref{fig:users_national}. We can see that this follows power law behaviour~\cite{tavares2013scaling}. That is, one can see that the number of posts by users drop off very quickly (a power relationship) as we increase the number of top posting users in our set. The number of weekly posts is shown in Fig \ref{fig:users_weekly}. We further highlight the length (number of words) of the social media posts in Fig \ref{fig:number_of_words} With this data, we now can focus on extracting topic patterns and analysing topics. 

\section{Topic extraction and analysis of discussions}
Preliminary experiments with an unsupervised topic model built on \textit{TweetsPublic} produced topics that could not be deciphered. Furthermore, the model poorly segmented the data into representative clusters.  In work by \cite{ghosh2017temporal,gallagher2017anchored}, a supervised topic modelling approach was used to seed the topic model with a manually curated set of words to guide the formulation of topics. Our approach aims to test whether topics created from a model trained on \textit{TweetsOfficial} and supervised by curated seed words can be used to supervise the training of a model trained on \textit{TweetsPublic} to identify associated topics in the larger, noisier corpus. If the \textit{TweetsOfficial}  model can effectively supervise the \textit{TweetsPublic} model to find topics that are associated with \textit{TweetsOfficial} topics, it will provide a reflection of the public response to the different topics of information propagated by government. The models were trained with 20 different topics and trained for 100 iterations. The models were used to label microblogs with a topic label to conduct analysis on the topics produced.

\begin{table}[t]
\caption{Seed Words }
\label{tbl:seed_words}
\centering
\tiny
\resizebox{0.7\columnwidth}{!}{%
\begin{tabular}{|l|l|}
\hline
\multicolumn{2}{|c|}{\textbf{Seed Words}} \\ \hline
movement,travel & \begin{tabular}[c]{@{}l@{}}coughing,nose,\\ touching,avoid,droplets\end{tabular} \\ \hline
\begin{tabular}[c]{@{}l@{}}death,cases,recovered,\\ recoveries\end{tabular} & kids,children,school \\ \hline
\begin{tabular}[c]{@{}l@{}}testing,tests,screening,\\ screen,swab\end{tabular} & economy,investment \\ \hline
\begin{tabular}[c]{@{}l@{}}alcohol,wine,\\ beer,drinking\end{tabular} & retrenched,lost,jobs \\ \hline
\begin{tabular}[c]{@{}l@{}}smoking,cigarettes,\\ cigarettes,smoke\end{tabular} & \begin{tabular}[c]{@{}l@{}}government,president,\\ minister,command\end{tabular} \\ \hline
\begin{tabular}[c]{@{}l@{}}home,distancing,\\ lockdown\end{tabular} & \begin{tabular}[c]{@{}l@{}}doctor,ppe,masks,\\ nurse,healthcare,hospital\end{tabular} \\ \hline
fake,news &  \\ \hline
\end{tabular}%
}

\end{table}

\subsection{Highlighted Topics}
The topics produced by the \textit{TweetsOfficial} and \textit{TweetsPublic} model were very similar because to the effects of the supervised approach taken. The topics in Table~\ref{tbl:topics} provide a general overview of COVID-19 themes that were central to the engagements during the observed period. The keywords used were similar for both models thus they are illustrated together in one column for brevity. The models produced topics directly related to COVID-19 and healthcare; the officials from Table~\ref{tab:accounts_tracked}; the imposed national lockdown regulations; and the policies applied by government for the lockdown including alcohol and cigarette bans, school closure, job loss mitigation and hygiene guidelines.

\begin{table}[t]
\caption{Highlighted topics }
\label{tbl:topics}
\centering
\tiny
\resizebox{\columnwidth}{!}{%
\begin{tabular}{|l|l|l|}
\hline
\textbf{Topic} & \textbf{Keywords} & \textbf{Label} \\ \hline
\textbf{1} & \begin{tabular}[c]{@{}l@{}}travel,movement,year,old,male,\\ female,travelled,relocation,italy,switzerland\end{tabular} & Travel \\ \hline
\textbf{2} & \begin{tabular}[c]{@{}l@{}}cases,recoveries,death,confirmed\\ ,total,number,deaths,today,recovered,related\end{tabular} & \begin{tabular}[c]{@{}l@{}}Case \\ Reports\end{tabular} \\ \hline
\textbf{3} & \begin{tabular}[c]{@{}l@{}}testing,tests,screening,conducted,\\ eligible,listed,feeling,flattenthecurve,hotline,sick\end{tabular} & \begin{tabular}[c]{@{}l@{}}Testing/\\ Screening\end{tabular} \\ \hline
\textbf{4} & \begin{tabular}[c]{@{}l@{}}alcohol,ly,bit,spreadthefacts,\\ guide,wash,vvnfkf,step,dry,ih\end{tabular} & \begin{tabular}[c]{@{}l@{}}Alcohol/\\ Hygience\end{tabular} \\ \hline
\textbf{5} & \begin{tabular}[c]{@{}l@{}}smoking,smoke,cigarettes,ukuthi,\\ abantu,lesifo,bakithi,ukhozi\_fm,uma,ngoba\end{tabular} & Cigarettes \\ \hline
\textbf{6} & \begin{tabular}[c]{@{}l@{}}lockdown,home,distancing,social,\\ essential,hygiene,stay,groceries,grants,graphic\end{tabular} & \begin{tabular}[c]{@{}l@{}}Lockdown/\\ Distancing\end{tabular} \\ \hline
\textbf{7} & \begin{tabular}[c]{@{}l@{}}avoid,touching,droplets,coughing,\\ nose,markets,tv,pscp,spoiled,stray\end{tabular} & \begin{tabular}[c]{@{}l@{}}COVID-19\\ Info\end{tabular} \\ \hline
\textbf{8} & \begin{tabular}[c]{@{}l@{}}children,school,earn,tax,reprieve,\\ salary,option,considering,uif,provisions\end{tabular} & \begin{tabular}[c]{@{}l@{}}Schools/\\ Jobs\end{tabular} \\ \hline
\textbf{11} & \begin{tabular}[c]{@{}l@{}}minister,president,mkhize,zweli,\\ ramaphosa,command,dr,cyril,\\ health,cyrilramaphosa\end{tabular} & \begin{tabular}[c]{@{}l@{}}Officials/\\ Command \\ Council\end{tabular} \\ \hline
\textbf{12} & \begin{tabular}[c]{@{}l@{}}hospital,healthcare,masks,\\ nurses,doctor,ppe,doctors,\\ nurse,collecting,workers\end{tabular} & \begin{tabular}[c]{@{}l@{}}PPE/\\ Healthcare\end{tabular} \\ \hline
\textbf{13} & \begin{tabular}[c]{@{}l@{}}fake,news,smart,kind,\\ stayathome,help,spread,ps,message,deal\end{tabular} & Fake News \\ \hline
\end{tabular}%
}

\end{table}

\subsection{Topic Timelines}
We use topic timelines to study the temporal properties of themes present in topics. Since the topics from both models are aligned in context, the objective is to analyse spikes in volume to understand the interaction between government and the public. Key dates such as lockdowns, and important pandemic notifications are highlighted to associate spikes in volume with events that transpired.
 
\begin{figure}[t]
    \centering
    \includegraphics[width=\columnwidth]{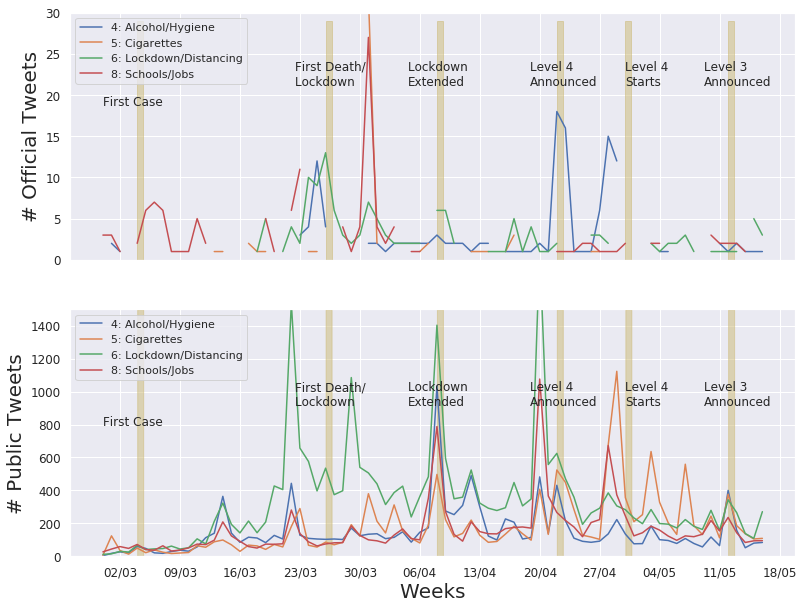}
    \caption{Official vs Public Topics on Lockdown Related Issues}
    \label{fig:alc_timeline}
\end{figure}

Fig~\ref{fig:alc_timeline} illustrates topic timelines for lockdown-related topics. The lockdown topic is consistently discussed with spikes in volume occurring near key dates that coincide with official communication and that of the public. Official communication spikes in volume at the start of lockdown but received limited coverage elsewhere. The cigarette topic spiked in public microblogs near certain dates, most notably in the period between level 4 being announced and being active because government changed their decision to allow cigarette sales in level 4 lockdown in that period. Official microblogs do not appear to focus on this topic even though this was a stark trend in public microblogs. Spikes were observed in the other topics aligned to themes related to lockdowns.

\begin{figure}[t]
    \centering
    \includegraphics[width=\columnwidth]{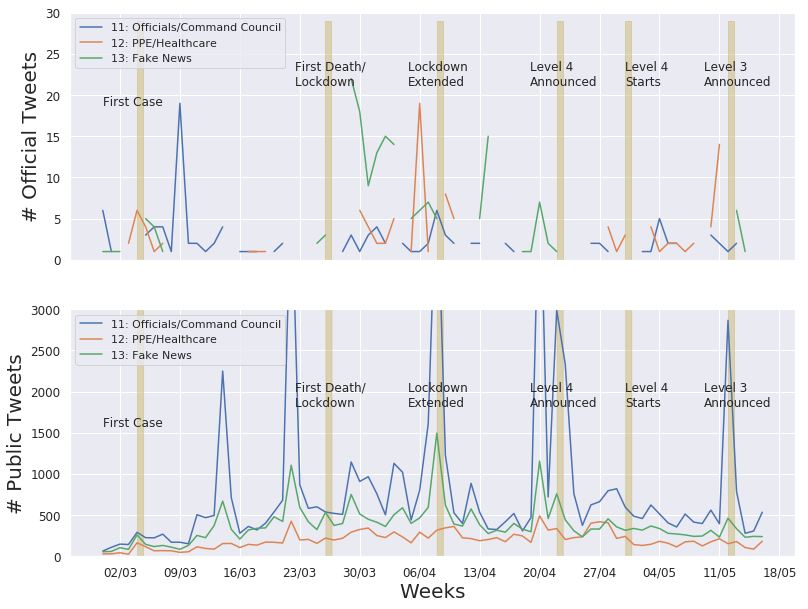}
    \caption{Official vs Public Topics on Other Issues}
    \label{fig:other_timeline}
\end{figure}

Fig~\ref{fig:other_timeline} illustrates topic timelines for topics related to public officials, personal protective equipment (PPE) and fake news. Public engagement towards key officials consistently occurred throughout the time period of the study, with peaks in volume near certain time periods. The spike in volume before the announcement of level 4 lockdown can be attributed to the public’s eagerness for the lockdown to be eased. The PPE/healthcare topic receives low volumes of attention but is consistently present in the public discourse. The fake news topic peaks alongside the public official communication trends around similar topics   

\subsection{Topic Similarity}
As a result of the supervised topic modelling approach, the \textit{TweetsOfficial} and \textit{TweetsPublic} models had topics described by very similar keywords. We use topic similarity heatmaps to analyse the syntactic similarity between topics from different models. The method consists of creating sub-corpora for each topic from each model, train a vectorizer on the union of a pair of sub-corpora, one from each model. Thereafter, a heatmap illustrates the pairwise cosine similarity of two models.  In this study, topic similarity heatmaps are used to understand the similarity between the official and public microblogs of the same topic. 

\begin{figure}[t]
    \centering
    \includegraphics[width=\columnwidth]{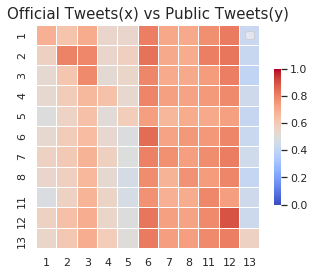}
    \caption{Topic similarity heatmap}
    \label{fig:topic_sim_heatmap}
\end{figure}

Fig~\ref{fig:topic_sim_heatmap} illustrates the topic similarity heatmap for the official and public topics. The strong similarity seen on the diagonal of the heatmap indicates that related topics from each model have a strong syntactic similarity to each other. The ’heat’ seen in most of the heatmap is a result of the data covering a limited domain of public engagements with government officials related to the pandemic. The very high similarity seen for topics 2 (Case Reports), 3(Testing/Screening) and 12 (PPE/Healthcare) are likely due to these topics being informative in nature resulting in the scope for public opinion being limited. Topic 6, 7, 8, 11 and 12 (see Table~\ref{tbl:topics} for descriptions) share common vocabulary with majority of the other topics. This is expected since most of these topics are directly related to the government’s response and the consequences on the public school closure and job loss because of the pandemic.

\section{Conclusion}
The topics produced highlight central themes in the discourse between government officials and the public. The topics address themes such as lockdown and the restrictions put in place, general information on COVID-19, testing reports and case reports, references to government officials central to the response and fake news, discourse on PPE and healthcare, and key issues like school closure and job losses. The topic timelines indicated that the public attempted to engage more closely to key dates in the observed period. Peaks in volume of topics provided insights into when certain topics received attention, most notably was the spike in the cigarette theme when the government changed their decision to lift the cigarette ban allow cigarette sales.

The high syntactic similarity seen between related topics from each model indicates that the supervised topic modelling approach taken can be used to find targeted insights from a larger noisier data set by seeding the training process with a smaller and more coherent data set that relates to the domain being studied. The application of this approach showed informative insights about the interactions between government officials and the public. Furthermore, the topics modelled in this study can be used in future studies related to supervised tasks aimed at stronger citizen engagement on social media.

\section{Acknowledgements}

We would like to thank the reviewers for their feedback, which made improved this work. We also want to acknowledge Herkulaas MVE Combrink for proof reading .This work is supported by the UP ABSA Chair of Data Science.

\bibliographystyle{./bibliography/IEEEtran}
\bibliography{covid19za-media.bib}
\end{document}